\documentclass{aastex}

\slugcomment{Submitted to AJ}

\shorttitle{Chamaeleon I: Near-Infrared Survey}
\shortauthors{GOMEZ AND KENYON}

\begin{document}


\title{A Near-Infrared Imaging Survey of the Chamaeleon I Dark Cloud}


\author{Mercedes G\'omez\altaffilmark{1}}
\affil{Observatorio Astron\'omico de C\'ordoba, Laprida 854, 5000 C\'ordoba, Argentina}
\email{mercedes@oac.uncor.edu}

\and

\author{ Scott J. Kenyon\altaffilmark{1}}
\affil{Smithsonian Astrophysical Observatory, 60 Garden Street, Cambridge MA 02138}
\email{skenyon@cfa.harvard.edu}


\altaffiltext{1}{Visiting Astronomer, Cerro Tololo Inter-American Observatory.
CTIO is operated by AURA, Inc.\ under contract to the National Science
Foundation.}


\begin{abstract}

We describe a near-infrared imaging survey covering $\sim$ 1 deg$^2$ 
of the Chamaeleon I dark cloud.  The survey is complete for $K$ $<$ 15.0, 
$H$ $<$ 16.0, and $J$ $<$ 16.5, roughly two magnitudes more sensitive
than previous large scale surveys. We use the large number of background stars
detected to derive an accurate near-infrared extinction law for the
cloud and select new candidate members with near-infrared
color excesses.  We list $\sim$ 100 candidates of the 
cloud with $K$ $\ge$ 12.0, based on their positions in the $J-H$, 
$H-K$ color-color diagram.  These new stars have low luminosities 
($K$ $\sim$ 12 -- 16, $H-K$ $\gtrsim$ 0.5 -- 1.5) and may have masses 
close to or even below the hydrogen burning limit.

\end{abstract}


\keywords{ISM: individual (Chamaeleon I) --- ISM: dust, extinction ---
star: formation --- stars: low-mass stars, brown dwarfs --- star: pre-main sequence}


\section{Introduction}

The Chamaeleon I dark cloud ($\alpha$ $\sim$ 11$^{\rm h}$, $\delta$
$\sim$ $-$77$^{\rm o}$)
is an active stellar nursery with 150 or more known young stars 
\citep{sch91,gast92,law96,cam98,com99}.
Located at a distance of $\sim$ 160 pc \citep{whi97},
it has a relatively small angular size \citep[$\sim$ 3 deg$^{\rm 2}$]{bou98}.
Because of its proximity and moderate extension on the sky the
Chamaeleon I star-forming region is an attractive place to attempt 
to measure the initial mass function (IMF), particularly at low masses
(i.e., for masses \hbox{$\le$ 0.1 M$_{\odot}$).}

Here we report results of a $JHK$ imaging survey of the Cha I dark cloud,
which covers an area of $\sim$ 1 deg${\rm ^2}$ and is complete for 
$K$ $<$ 15.0, $H$ $<$ 16.0, and $J$ $<$ 16.5.  Our survey
complements the $IJK_{\rm s}$ observations of DENIS (Deep Near 
Infrared Southern Sky Survey, ESO, 1m tel, \citet{epc97}) 
described by \citet{cam98}.  The DENIS survey covers a larger
area, 1.5$\rm{^o}$ $\times$ 3$\rm{^o}$, of the cloud; but, with
a sensitivity of $K_s$ $<$ 13.5, it is less sensitive to the lowest 
mass cloud members. \citet{cam98} proposed 54 new candidate 
young stars, based on their locations in the $I-J$/$J-K$ diagram. 
In addition our survey area comprises two areas,
in the the northern region, recently observed by
\citet{per99} and \citet{oas99}. The data reported by these
authors are $\sim$ 2 magnitudes deeper than ours but cover 
modest areas on the cloud ($\sim$ 3$'$ $\times$ 3$'$, \citet{per99};
$\sim$ 6$'$ $\times$ 6$'$, \citet{oas99}) centered close to the high-velocity
C$^{18}$O bipolar outflow previously detected by \citet{mat89}. 
\citet{per99} identified a new Class I 
source of the cloud. \citet{oas99} proposed 9 new very low luminosity
members of the cloud. 

Our survey recovers 40 of these DENIS candidates (lying on common areas to 
both surveys) and the stars with $K$ $<$ 15.5 detected by \citet{per99} and
\citet{oas99}. In addition the present observation yields $\sim$ 100 new candidates with
$K$ $\ge$ 12 and near-infrared excess emission in the $J-H$,$H-K$ color-color diagram.
The low luminosities of these candidates suggests they may be very low mass cloud
members, with masses close to or below the hydrogen burning limit.

We describe our observations, data reduction, and analysis in \S2.
The astrometry and photometry were obtained for $\sim$ 11,090
sources at $JHK$ detected in our survey region. 
In \S 3 we derive a reliable extinction curve for
background stars and use this curve to propose $\sim$ 100 new
potential young stellar objects based on their location in the $J-H$/$H-K$ diagram
in \S 4.  We conclude with a brief summary in \S 5.

\section{Observations and Data Analysis}

We obtained $JHK$ imaging data for the Chamaeleon I dark cloud and three 
relatively unreddened control fields on 13--16 February 1995, 8--11 March 1996,
16--19 April 1997, and 1--2 November 1998 with CIRIM (the Cerro Tololo Infrared Imager) at 
the CTIO 1.5m telescope.  The CIRIM uses a 256 $\times$ 256 HgCdTe NICMOS 3 
array, which provides a field of $\sim$ 4.9$'$ $\times$ 4.9$'$ with
a plate scale of 1.16$''$ per pixel. We covered an area of 
\hbox{$\sim$ 0.65$^o$ $\times$ 1.5$^o$} on the cloud on a rectangular grid,
with 1$'$ overlap between adjacent frames. We acquired two 6 $\times$ 5 
sec exposures for each field, shifted by 20$''$. Figure 1 shows the extent 
of the survey relative to the H$_2$CO contour map measured by \citet{toma85} 
and the distribution of previously known young members of the cloud 
\citep{sch91,gast92,law96,cam98,com99}.  Cha I has a remarkable low-mass star formation 
activity localized mainly in three groups.  These small clusters are situated 
near three reflection nebulae: Ced 112 (HD 97300), Ced 110, and Ced 111 
(HD 97048), from North to South.

Our three control fields, covering an area of $\sim$ 0.05 deg${\rm ^2}$
and selected from a visual inspection of the ESO Red Sky Survey prints,
are close to the survey region and relatively free from significant 
optical extinction. These off-cloud regions lie outside the limits of 
Figure 1, $\sim$ 1.5--2 deg away from the H$_2$CO boundaries, one to 
the North and the other two to the East and West, respectively.
Table 1 gives positions and total areas covered by each of the 
off-cloud regions.

We processed the data using standard techniques with the software 
package IRAF\footnote{IRAF is distributed by the National Optical 
Astronomy Observatory, which is operated by the Association of 
Universities for Research in Astronomy, Inc. under contract to the
National Science Foundation.}.  From each program frame, we subtracted 
an average dark frame, divided by a normalized flat-field frame appropriate
for each filter, and then subtracted a flattened sky frame. The dark 
frame is the average of $\sim$ 40 individual dark images taken at 
the beginning and end of each night.  We constructed flat-field frames 
for each night by median-filtering all 6 $\times$ 5 sec frames taken 
in each filter. We also obtained a set of `` dome flats '' in each filter
before observing each night using an illuminated white spot in
the CTIO 1.5-m dome. Flat-fields constructed from `` dome flats '' and
`` sky flats '' were identical.  Sky frames were generated from a median-filtered
set of 20--30 flattened frames in each filter. In this case, we combined
individual object frames obtained close in time and in position to
the program frame. 

To produce a combined image from each dithered pair of images, we 
aligned the frames using the IRAF subroutines GEOMAP and GEOTRAN, 
added the co-aligned frames together, and trimmed the resulting 
image to remove bad pixels at the edges of the frames.  We selected 
the $K$ images as the reference frames and transformed the $J$ and 
$H$ images to the same pixel scale as the $K$ images. We used DAOFIND 
to locate stars 4$\sigma$ above the local background and added to the 
DAOFIND list all stellar objects missed by this routine, found by visual
inspection of each image. We then derived photometry for each image using
the APPHOT PHOT task, using a circular aperture with 5$''$ radius. This 
aperture size includes the total flux for the large majority of sources. 
However, in some cases, we used a smaller aperture (i.e. 2.5$''$) to 
avoid contamination from very close stars. Several iterations of this 
process produced an homogeneous set of data for the complete survey.
We detected $\sim$ 11,090 sources at $JHK$ in our survey region. 

To calibrate our photometry, we observed on each night a set of 10-15 
standards from \citet{eli82} and from the UKIRT faint $JHK$ standard 
stars list \citep{caha92}.  We estimate an uncertainty of $\pm$ 0.03 mag 
in our calibration. The standard stars were observed at a similar airmass
range as our target fields.  Airmass corrections for our data are smaller
than our photometric uncertainties. We used common stars on dithered pairs 
and measurements of duplicate stars in the overlapping regions of adjacent 
frames to estimate photometric uncertainties for the stars in our survey
region. Table~\ref{tbl-2} lists typical photometric uncertainties in each magnitude 
bin.  These differences reach 0.3 mag for $J$ $\sim$ 17, $H$ $\sim$ 16, 
and $K$ $\sim$ 15.  The 5$\sigma$ limiting magnitudes are $K$ $=$ 14.5, 
$H$ $=$ 15.5, and $J$ $=$ 16.0.

To make a check on the completeness of the survey, we constructed a list 
of sources from the three off-cloud regions (see Table 1).  Figure 2 shows 
the magnitude histograms of background stars as function of the $JHK$ 
magnitudes, The number of stars per bin increases monotonically up to 
the completeness limit and then turns over. The corresponding limits in 
each filter are: 16.5 at $J$, 16.0 at $H$, and 15.0 at $K$.

To compare our photometry with results for previous surveys, we observed 
$\sim$ 50 stars projected on Bok globule 2 in the Coalsack \citep{tap75}.
\citet{jon80} and \citet{eli83} previously reported photometry of these
stars and derived a comparison between the Mt. Stromlo/AAO photometry of 
\citet{jon80} and the CIT photometric system of \citet{eli82}.  
The common photometric data between \citet{jon80} and our survey, 
together with the transformations listed in \citet{eli83}, allow us 
to derive a transformation between the CIRIM and CIT photometric systems.  
Using the \citet{pres92} routine FITEXY, straight line fits to the 
common data yield:

\begin{equation}
\rm K_{CIRIM} = +0.03\pm0.02 ~+~ 0.99\pm0.01 \cdot K_{CIT}
\end{equation}

\begin{equation}
\rm (J-H)_{CIRIM} = -0.04\pm0.03 ~+~ 0.95\pm0.02 \cdot (J-H)_{CIT}
\end{equation}

\begin{equation}
\rm (H-K)_{CIRIM} = -0.02\pm0.02 ~+~ 1.01\pm0.02 \cdot (H-K)_{CIT}
\end{equation}

\noindent
The quoted uncertainties in these relations are the 1$\sigma$ errors 
from our fits to the Mt. Stromlo/AAO data and those from the \citet{eli83} 
fits, added in quadrature.  The differences between the 
CIRIM and CIT system are negligible for $K$ and $H-K$, in agreement
with results quoted in the CIRIM manual written by R. Elston and J.
Elias\footnote{Available 
at http://www.ctio.noao.edu/instruments/ir\_instruments/cirim/cirim.html}.
We thus assume that the natural CIRIM system is identical to the CIT
system for $K$ and $H-K$. Our derived color term for $J-H$ agrees with 
Elston \& Elias;  the CIRIM $J-H$ color is bluer than the CIT $J-H$ 
due to its different lens design.  We apply the color term 
to our $J-H$ data and thus quote colors in the CIT system.

To derive coordinates for our survey stars, we used WCSTool\footnote{
Available at ftp://cfa-ftp.harvard.edu/pub/gsc/WCSTools} 
\citep{min97}, a suite of programs to calculate a direct transformation 
between the coordinates of the image (x,y) and the sky coordinates 
($\alpha$,$\delta$). We measured transformation coefficients adopting 
matches between program stars and stars in the U.S.  Naval Observatory 
SA1.O Catalogue\footnote{The U.S. Naval Observatory SA1.0 Catalogue is
distributed by the U.S. Naval Observatory, Washington DC.}. These matches 
provided a good transformation for $\sim$ 90\% of the frames.  For the 
rest of the frames, usually corresponding to the most obscured regions, 
we found few or no matches between our frames and stars in the catalog.
In these cases, we used stars in common with adjacent frames with good
coordinates to obtain coordinates for stars on unmatched frames. When 
an individual source was detected in more than one filter we adopted 
the average coordinate.

To verify the coordinates, we measured coordinate differences between
stars on adjacent frames.  Many stars have two good coordinates;
others have as many as four.  After visually inspecting the duplicate
stars and rejecting stars too close to the edges of frames, we averaged
coordinates of the remaining duplicates and derived 1$\sigma$ errors. 
Finally, we compared coordinates derived from the Digitized Sky 
Survey\footnote{Based on photographic data obtained using The UK Schmidt 
Telescope. The UK Schmidt Telescope was operated by the Royal Observatory 
Edinburgh, with funding from the UK Science and Engineering Research 
Council, until 1988 June, and thereafter by the Anglo-Australian 
Observatory. Original plate material is copyright (c) the Royal
Observatory Edinburgh and the Anglo-Australian Observatory. The 
plates were processed into the present compressed digital form 
with their permission. The Digitized Sky Survey was produced at 
the Space Telescope Science Institute under US Government grant NAG W-2166.
Copyright (c) 1993, 1994, Association of Universities of Research in 
Astronomy, Inc. All right reserved.} (DSS) plates. We estimate an
average uncertainty in our positions of 1$''$. These errors are as 
large as 2$''$ for sources in the most obscured regions of the cloud.

Having verified the quality of the coordinates and photometry for our
survey, we search for pre-main sequence stars with near-infrared excess 
emission in the cloud using the near-infrared color-color diagram.  To identify 
these stars, we derive a reliable extinction curve for background stars 
in \S3 and then consider the near-infrared color-color diagram in \S4.




\section{The Cha I Near-Infrared Reddening Law}

To derive the near-infrared reddening law for the Cha I dark cloud, we follow 
\citet{ke98a}, who developed a generalized photometric 
technique to compare the colors of reddened stars with the colors of
nearby, `` unreddened '' comparison stars.  \citet{ke98a} assume that
the stellar population behind the cloud is identical to the stellar
population in off-fields several degrees away.  They derive the $J-H$ 
and $H-K$ color excesses for each on-cloud source relative to every
off-cloud source and then compute the average and median color excesses 
for each on-cloud source.  The probable error of the average color excess
is the sum in quadrature of the errors of the on-cloud and off-cloud
colors. For the median color excess, the probable error is the 
inter-quartile range.  We divide the Cha I sources into a Complete Sample
containing all stars with $K < 14$ and a Restricted Sample, where we remove
known young stars in the cloud from the Complete Sample.  The sample of 
known young stars includes 94 sources (lying on our survey region)
from the literature \citep{sch91,gast92,law96,cam98,com99}.
This sample probably is not complete; however, any incompleteness
does not affect our analysis significantly.

Figure 3 shows average color excesses of Cha I stars for two 
K $\le$ 14 samples, the Complete Sample and the Restricted Sample.  
The colors have been transformed to the CIT system using equations 1--3. 
The color excesses are highly correlated, with a Spearman rank-order 
correlation coefficient of $r_s$ $=$ 0.65 for both samples.  The
probability for no correlation between the two color excesses is
formally zero according to the Spearman rank-order test. Straight 
line fits to color excess measurements with $E_{H-K} \le 2.0$ yield
$E_{J-H}/E_{H-K}$ $=$ 1.63 $\pm$ 0.03 for the Complete Sample and 
$E_{J-H}/E_{H-K}$ $=$ 1.76 $\pm$ 0.02 for the Restricted Sample.
We derive identical slopes using median color excesses for each 
source instead of the average color excesses.  

Our extinction results for the Restricted Sample are identical to the 
\citet{bebr88} near-infrared extinction law, $E_{J-H}/E_{H-K}$ $=$ 1.75
transformed to the CIT photometric system.
The Cha I results differ from the near-infrared extinction law for $\rho$ Oph, 
$E_{J-H}/E_{H-K}$ $=$ 1.57 $\pm$ 0.03 \citep[for example]{ke98a,eli78} 
at more than the 3$\sigma$ level, and are inconsistent with results from
the \citet{he95}  survey of luminous southern stars, $E_{J-H}/E_{H-K}$ 
$=$ 1.47 $\pm$ 0.06, at roughly the 3$\sigma$ level.  

Results derived for the Complete Sample demonstrate that pre-main 
sequence stars with near-infrared excesses skew the extinction law to smaller 
values (see also \citet{ke98a} and references therein). Figure 3 
shows many previously identified sources with apparent near-infrared excesses.
These sources probably skew our results for the reddening law derived
from the Restricted Sample, although this effect should be small. 
\citet{ke98a} described a method, using the `` reddening probability
distribution '', to correct the reddening law for previously unidentified
pre-main sequence stars and to derive a sample of candidate pre-main
sequence stars with near-infrared excess emission. We now apply this method 
to the Restricted Sample of Cha I sources.

\citet{ke98a} defined the reddening probability distribution,
$\rho(E_x,E_y)$, as the probability of measuring a pair of color
excesses, $E_x$ and $E_y$, where $x$ and $y$ are color indices and
$N = \int \rho(E_x,E_y) ~ dE_x dE_y$ is the number of reddening 
measurements. The number, $N$, is also the number of on-field stars 
in the sample.  The density function, $\rho(E_x,E_y)$, depends on 
the distributions of colors in the on-field and the off-field.  If
$\rho_{off} (x,y)$ is the off-field color distribution and 
$\rho_{on} (x,y)$ as the on-field color distribution, the reddening 
probability distribution is:

\begin{equation}
\rho_{i,j}(E_x,E_y) = \sum_{k_1=1}^{N_1} \sum_{k_2=k_{min}}^{k_{max}} ~ \rho_{k_1,k_2,off}(x,y) ~  \rho_{k_3,k_4,on}(x,y)
\end{equation}
 
\noindent
where $k_3 = i+k_1-i_0$ and $k_4 = j+k_2-j_0$.  The integers $i_0$ and
$j_0$ measure the zero point offsets of the color grids; indices $i$
and $j$ span the full range of color indices $x$ and $y$.  The color
distributions use a the kernel density estimator:
 
\begin{equation}
\rho(x,y) = \frac{1}{h^2}\sum_{i=1}^{N} K(x,x_i,y,y_i) ~ ,
\end{equation}
 
\noindent
where $h$ is the smoothing parameter and K is the kernel \citep{sil86}.
We follow \citet{ke98a} and adopt
 
\begin{equation}
K({\bf x}) = \left\{ \begin{array}{l l l}
                \frac{4}{\pi} (1 ~-~ {\bf x^T x})^3 & \hspace{5mm} & {\rm if} ~
{\bf x^T x} < 1 \\
                0                             &              & {\rm otherwise} \\
        \end{array}
        \right.
\end{equation}
 
\noindent
for fast computation ({\bf x} = ($x,~x_i,~y,~y_i$)).
We adopt $h = 0.2$, which is roughly twice the 1$\sigma$ error of
our photometry at the $K = 14$ mag survey limit.  Smaller values 
for $h$ produce noisy grids; larger values are inconsistent with 
the photometric errors.

Figure 4 shows the color distribution functions for the off-field 
and the Complete Sample (CS). The off-field density in the left panel
has a weak maximum at the color expected for G-type giants and a 
sharp maximum at the color expected for K-type dwarfs. There
is a plateau consistent with colors for M-type dwarfs and
an extension of the dwarf locus towards the near-infrared colors of 
K- and M-type giant stars. The Cha I color distribution in the 
right panel peaks close to the off-field maximum but is much 
more extended.  The long axis of the contours roughly follows
the reddening line derived above, and there is a weak extension of 
sources with near-infrared excess emission at $H-K$ $=$ 0.75 and $J-H$ $=$ 1.75.

Figure 5 shows the reddening probability functions constructed
from the color distributions for the Complete Sample (CS) and
the Restricted Sample (RS).  We derive the slope of the
reddening law which follows the major axis of each contour
in these diagrams using the technique outlined in \citep[equations 12--15]{ke98a}.
This analysis yields 
$E_{J-H}/E_{H-K}$ $=$ 1.77 $\pm$ 0.03 for the Complete Sample and 
$E_{J-H}/E_{H-K}$ $=$ 1.80 $\pm$ 0.03 for the Restricted Sample.
These results are consistent with the reddening law derived in
Figure 3 given the 1$\sigma$ errors of each fit.

\section{New Candidate Pre-Main Sequence Stars}

Figure 6 shows the $H-K$/$J-H$ diagram for our complete survey region.
As in Figures 3--5, we display stars with \hbox{$K$ $<$ 14.0},
where the photometric errors are $\lesssim$ 0.1 mag. The solid 
line corresponds to the main sequence locus \citep{bebr88}; 
the dashed lines, parallel to the reddening vector derived in \S3, 
define the reddening band extending from the main sequence.

We distinguish two groups of sources in Figure 6: a) stars that lie along
or follow the reddening band direction, and b) sources that show near-infrared
excess, located to the right of the reddening band. The first group, comprising
the majority of the sources, show negligible near-infrared color excess,
and can be dereddened to lie close to the standard dwarf sequence.  Most of 
these sources are probably behind the cloud\footnote{The WTTS that belong to
the cloud and that, as a class of young stellar objects, has no significant near-infrared
excess may be included in this group.}.  The color excess of the second sample 
cannot be attributed to the reddening by dusty material in the cloud. Several 
authors have found that circumstellar material (disks + infalling envelopes) surrounding the
young central star can produce these color excess at the near-infrared wavelengths
\citep{laad92,keha95}.  Formally speaking we have detected \hbox{$\sim$ 300} sources
lying to the right of the reddening band in our survey with \hbox{$K <$ 14.0.}
Roughly 1/3 are known members of the cloud and DENIS candidates from 
\citet{cam98}.  The other 2/3 are new detections.
In \hbox{Figure 6} we have used starred symbols to indicate new
candidates with significant near-infrared excess (i.e., $H-K$ $\gtrsim$ 0.5).
This group comprises $\sim$ 50 objects.
Table 3  gives coordinates and
magnitudes for these stars. The rest of the sources either lie very close to the
reddening line or to the bulge defined by the background stars.

Finally, our survey has also produced a list of $\sim$ 50 faint objects
(i.e., $K$ $>$ 14 -- 16 ), with large near-infrared excess (i.e., $H-K$ $\gtrsim$ 0.8).
Figure 7 shows the location of these stars in the color-color diagram.
They are located well to the right of the main sequence locus. In this manner even
allowing typical photometric uncertainties for these magnitudes
(i.e., $\sim$ 0.3 mag for $K$ $\sim$ 15) the sources still have color excesses.
Table 4 gives coordinates and magnitudes for these stars. 
For these faint sources, we checked each of the images visually to avoid 
possible confusion
with cosmic ray events, bad pixels, plate artifacts, or contamination with 
close stars.

A few of the objects listed in the Table 3 and Table 4 were only detected in two of
the filters ($K$ and $H$) and therefore $J$ must be $\gtrsim$ 18.0. These are extremely
red objects and thus good candidate members of the cloud. Some of the objects
display extended or fuzzy images principally at $K$. Four
candidates were also detected by the ISOCAM survey of the Chamaeleon I dark cloud
and proposed as new low-mass candidate members based on their position on the
mid-infrared color-color diagram \citep{per00}. We also have in common two
stars with \citet{oas99} (see Table 3 and Table 4).                                    

Sources with near-infrared excess emission are either young stars in the cloud,
galactic sources such as planetary nebulae (e.g., \citet{gug98}), or galaxies.
Extragalactic source counts predict 30--60 galaxies in an area similar
in size to our survey region for $K \le$ 14 (e.g., \citet{szo98,vai00} and the
references therein).  The near-infrared extinction towards Cha I is significant
\citep{cam97};
we thus expect $\sim$ 30\% to 50\% fewer galaxies in our survey compared to
unobscured regions.  Galaxies also tend to have smaller near-infrared excesses
compared to young stars.  Most galaxies have colors similar to red main
sequence stars, $J-H$ $\approx$ 0.6--0.8 and $H-K$ $\approx$ 0.25--0.45,
with modest near-infrared excesses of 0.2--0.3 mag in H--K (e.g.,
\citet{fro85,imp86}).
Unpublished 2MASS data support this conclusion for nearby galaxies (see
\citet{jar98}\footnote{Available also at
http://spider.ipac.caltech.edu/staff/tchester/2mass/analysis/galaxies/colors/}).
Our survey is not deep enough to detect more distant galaxies which could
have redder near-infrared colors than measured for the nearby galaxy samples.
 
Several types of galactic sources, such as carbon stars and planetary nebulae,
can also lie outside the reddening band in the near-infrared color-color diagram.
The surface densities of these objects towards Cha I should be small,
$\le$ 5--10 deg$^{-1}$ for carbon stars and $\le$ 1--2 deg$^{-1}$ for
planetary nebulae (e.g., \citet{gug98,orma94}).
 
The spatial distribution of the near-infrared excess sources provides the best
evidence that these objects are pre-main sequence stars.  Figure 8 compares
the spatial distribution of the new candidate objects with the positions of
previously known young stars \citep{sch91,gast92,law96,cam98,com99}
and the H$_2$CO contour maps \citep{toma85}.  The new candidates tend
to cluster around Ced 112 (HD 97300), Ced 110, and Ced 111 (HD 97048)
in a manner similar to the previously known members of the cloud (see
Figure 1).  We expect extragalactic and other galactic sources such as
carbon stars and planetary nebulae to avoid regions of high obscuration
and see no evidence for this tendency in our candidates.                      

Assuming an average inter-cloud extinction of A$_V$ $\sim$ 5
\citep[i.e., A$_K$ $\sim$ 0.5]{cam97,cam99}, a main-sequence
star at the H burning limit and at the distance of the Cha I cloud 
\citep[160 pc]{whi97}, would have $K$ $\sim$ 15.5 -- 16.0 \citep{hemc93}.
Low mass members of the Chamaeleon I dark cloud must be much younger,
still contracting toward the main sequence, and thus significantly more luminous than
main sequence stars of 0.1 M$_{\odot}$. Our imaging survey is complete to $K$ $\sim$ 15.0 and
sensitive down to 16.5. The objects detected in this survey (with $K$ $\sim$ 12 -- 16,
$H-K$ $\gtrsim$ 0.5 -- 1.5) are among the lowest luminosity and, presumably, lowest mass members
of the cloud. Some of them are near or, probably, below the H burning limit.

Some of our candidates (with $K$ $\sim$ 12 -- 14) have optical counterparts on the DSS plates.
We crudely estimate \hbox{V $\sim$ 17 --  22} for this subgroup of candidates. Moderate
resolution optical spectra would allow a search for Li I 6707 absorption, an indicator
of youth, as well as other atomic (such as H$\alpha$) and many forbidden emission lines
usually present in the spectra of young stellar objects \citep[for example]{ke98b}.
To investigate further the nature of
the rest of our candidate objects we need additional 10 $\mu$m photometry and/or near-infrared
spectroscopic observations. In particular, near-infrared spectra would allow us to place
the stars in the HR diagram and thus, adopting recently developed pre-main sequence tracks
\citep[for example]{dama97,past99}, determine masses and ages for the individual
sources. Mass determinations are needed to estimate the lower end of the IMF. Present
determinations of the IMF in the Cha I cloud are limited to 0.3 M$_{\odot}$ and for
masses $\lesssim$ 0.6 M$_{\odot}$ these determinations are uncertain
\citep{app83,law96}.  Age estimations will provide additional support
to the concept of a coeval star-formation process in this cloud
\citep{gast92}.

\section{Summary}

We have used the large number of background stars detected in our survey
region to derived a reliable near-infrared extinction law for the Chamaeleon I 
cloud. Our analysis yields $E_{J-H}/E_{H-K}$ $=$ 1.80 $\pm$ 0.03 for the 
Restricted Sample (i.e., eliminating previously known pre-main sequences
stars).  This result differs with the \citet{ke98a}'s determination for
the $\rho$ Oph cloud ($E_{J-H}/E_{H-K}$ $=$ 1.57 $\pm$ 0.03) at more than
3$\sigma$ level. Based on preliminary results for other two cloud (Taurus
and IC 348), in addition to their determination for $\rho$ Oph,
\citet{ke98a} suggest that a real variation in the near-infrared reddening
law occurs from region to region. Our results for Cha I provide additional
support for this variation. 

The ratio of total to selective extinction, R$_V$, seems also to change for
different clouds (see \citet{ke98a} and the references therein).  In addition,
R$_V$ varies across the Cha I dark cloud
(from R$_V$ $\sim$ 3 to R$_V$ $\sim$ 5.5; \citet{whi97}, see also
\citet{hyl82}).  For the $\rho$ Oph, \citet{vrb93} obtained a mean value of
R$_V$ $\sim$ 4, although individual determinations vary within a
range similar to that found for the Cha I dark cloud.
\citet{hay99} recently found that the
A$_V$ -- $N(CO)$ relation also varies among different star-forming clouds and
suggest that A$_V$/$N(CO)$ may also change across the Chamaeleon I dark cloud.
These results indicate different dust-grain properties, in particular
grain-sizes, for different nearby star-forming regions and caution against 
using the `` universal '' reddening law to derive extinction estimates for
young stars in star-forming regions. 
 
Previous near-infrared observations of the cloud focussed on IRAS-selected sources
\citep{ass90,whi91,pru91} or already known
and candidate young stellar objects detected at optical and X-ray wavelengths
\citep{gast92,har93,law96,com99}. These investigations were sensitive to
objects with K $\sim$ 11 and limited to small areas around the target objects. 
Present imaging near-infrared surveys of the cloud have overcome these  
limitations (both in sensitivity and spatial coverage) and greatly increased the number of
potential members of the cloud. \citep{per99} detected one new Class I source of
the cloud with K $\sim$ 13. \citet{oas99} proposed 9 new candidates with 
K $=$ 13 -- 16.  \citep{cam98} and this paper carry out large scale
near-infrared survey on the cloud.  \citep{cam98} proposed 54 new candidates
members with K $\lesssim$ 13. This paper presents an additional set of
$\sim$ 100 new stars with K $\sim$ 12 -- 16.
In addition to these candidates selected on basis of their near-infrared color
excesses, \citet{per00} have recently proposed a list of 74 new low mass
candidates detected by ISOCAM and selected on basis of their color excesses
at the mid-infrared wavelengths.              

These objects are among the lowest luminosity and, presumably, lowest
mass members of the cloud. Some of them are near or, probably, below the H burning limit.
A spectroscopic follow up of the optically visible candidates should provide additional
indications of their pre-main sequence nature. Optical and/or near-infrared spectral types
of our proposed candidates, in combination with recently developed pre-main
sequence evolutionary tracks, will allow us to determine masses and ages.
Reliable mass determinations are required to calculate the IMF, particular towards
the sub-solar and probably into the sub-stellar mass regime. Ages
determination will help to reconstruct the history of the star-formation process
of the cloud.

\acknowledgments

We are grateful to the CTIO staff, specially to M. Fern\'andez,
M. Hern\'andez, and P. Ugarte for assistance during the observing runs.
We also thank R. Elston and R. Probst for their help with CIRIM, D.
Mink for assistance with the WSCTools software and the anonymous referee
for suggestions that improved the manuscript. 
This research was partially
supported by the Scholarly Studies Program of the Smithsonian Institution and
the National Aeronautics and Space Administration (grants: NAGW-2919 and GO-06132.01.94A).
M.G. acknowledges support from the National Science Foundation through grants GF-1001-96 and
GF-1001-97 from the Association of Universities for Research in Astronomy, Inc., under NSF
cooperative agreement AST-8947990.

\clearpage



\newpage

\figcaption[fig1.ps]{The extent of our near-infrared survey relative to the
H$_2$CO contour maps from \citep{toma85} and the distribution of previously
known members of the cloud \citep[dots]{sch91,gast92,law96,cam98,com99}.
The large starred symbols indicate the positions of
Ced 112 (HD 97300), Ced 110, and Ced 111 (HD 97048) from North to South,
respectively.\label{fig1}}

\figcaption[fig2.ps]{Histogram distribution of magnitudes for background stars in our control fields.
The completeness limit in each band is indicated by the dashed vertical lines.\label{fig2}}

\figcaption[fig3.ps]{Average color-excess diagram for two samples of the Chamaeleon I stars
with K $\le$ 14.0. The left panel shows the Complete Sample (CS) and the right panel the
Restricted Sample (RS).
This second sample only includes near-infrared sources not known to be pre-main
sequence stars. \label{fig3}} 

\figcaption[fig4.ps]{Color density distribution for sources in the off-field region (left panel)
and for Cha I sources Complete Sample (CS, right panel). \label{fig4}} 

\figcaption[fig5.ps]{Reddening probability distribution for the Complete Sample
(CS) and the Restricted Sample (RS). The dot-dashed line indicates our best reddening
law; the dashed line shows a standard reddening law with $E_{J-H}/E_{H-K}$
$=$ 1.75. \label{fig5}}

\figcaption[fig6.ps]{Color-color diagram for near-infrared sources
with $K$ $<$ 14.0 in our survey region (see Figure 1).  The solid line indicates the main
sequence locus \citep{bebr88} and the two parallel lines define the reddening band,
extending from the main sequence. To define this band we used the reddening
vector ($E_{J-H}/E_{H-K}$ $=$ 1.80 $\pm$ 0.03) derived for the Restricted Sample (RS in Figure 5,
left panel).  Typical photometric errors for $K$ $\sim$
14.0 are displayed in the upper left corner.\label{fig6}}

\figcaption[fig7.ps]{Color-color diagram for near-infrared sources
with $K$ $>$ 14.0 in our survey region (see Figure 1).  The solid line indicates the main
sequence locus \citep{bebr88} and the two parallel lines define the reddening band,
extending from the main sequence. 
To define this band we used the reddening
vector ($E_{J-H}/E_{H-K}$ $=$ 1.80 $\pm$ 0.03) derived for the Restricted Sample
(RS in Figure 5, left panel).  Typical photometric errors for $K$ $\sim$
15.0 are displayed in the upper left corner.\label{fig7}}

\figcaption[fig8.ps]{Spatial Distribution of the new candidate young stellar objects 
(triangles) listed in the Table 3 and Table 4 in relation to the H$_2$CO contour maps \citep{toma85}
and the previously known members of the cloud
\citep[dots]{sch91,gast92,law96,cam98,com99}.
The large starred symbols indicate the position of Ced 112 (HD 97300), Ced 110, and Ced 111
(HD 97048) from North to South, respectively.\label{fig8}}





\clearpage

\begin{deluxetable}{cccc}
\tablecaption{Off-Cloud Regions. \label{tbl-1}}
\tablewidth{0pt}
\tablehead{
\colhead{Region} & \colhead{$\alpha$(2000.0)} & \colhead{$\delta$(2000.0)} & \colhead{Area} }
\startdata
I& 11 31 43& $-$77 38 22& 5$'$$\times$5$'$\\
II& 10 41 24& $-$77 38 35& 10$'$$\times$10$'$\\
III& 11 16 52& $-$77 08 42& 5$'$$\times$10$'$\\
 \enddata

\tablecomments{Units of right ascension are hours, minutes, and seconds, and
units of declination are degrees, arcminutes, and arcseconds.}




\end{deluxetable}

\clearpage

\begin{deluxetable}{cccc}
\tablecaption{Mean Photometric Errors. \label{tbl-2}}
\tablewidth{0pt}
\tablehead{
\colhead{Mag. Bin} & \colhead{$K$} & \colhead{$H$} & \colhead{$J$} }
\startdata
 8 -- 9  & 0.02 & 0.01 &  0.01 \\
 9 -- 10 & 0.03 & 0.02 &  0.02 \\
10 -- 11 & 0.03 & 0.02 &  0.02 \\
11 -- 12 & 0.03 & 0.03 &  0.03 \\
12 -- 13 & 0.04 & 0.03 &  0.03 \\
13 -- 14 & 0.07 & 0.04 &  0.03 \\
14 -- 15 & 0.12 & 0.08 &  0.05 \\
15 -- 16 & 0.26 & 0.12 &  0.07 \\
16 -- 17 & 0.46 & 0.20 &  0.09 \\
17 -- 18 &\nodata& 0.38 &  0.15 \\
\enddata

\end{deluxetable}

\begin{deluxetable}{cccccccl}
\tablecaption{Candidate Pre-Main Sequence Stars with $K$ $<$ 14.0. \label{tbl-3}}
\tablewidth{0pt}
\tablehead{
\colhead{Star}& \colhead{$\alpha$(2000.0)} & \colhead{$\delta$(2000.0)} & \colhead{$K$} &
\colhead{$H-K$} & \colhead{$J-H$} & \colhead{Id.}  & \colhead{Ref.} }
\startdata
1& 11 04  34.0& $-$78 01 26& 13.34&  0.68&  1.00& & \\
2& 11 04  41.5& $-$77 38 52& 13.64&  0.73&  1.09& &\\
3& 11 04  49.2& $-$78 04 46& 12.48&  0.75&  1.03& & \\
4& 11 04  59.5& $-$77 17 00& 12.59&  0.85&  0.84& & \\
5& 11 05  27.2& $-$77 33 09& 11.80&  2.99&  1.24& & \\
6& 11 05  43.0& $-$77 29 13& 13.79&  1.34&  1.74& & \\
7& 11 05  43.4& $-$77 21 22& 13.81&  0.83&  0.85& & \\
8& 11 05  48.8& $-$76 40 17& 13.94&  0.73&  0.55& & \\
9& 11 05  57.8& $-$76 40 56& 13.06&  0.84&  0.79& & \\
10& 11 06  08.8& $-$76 56 02& 13.56&  0.73&  0.91& & \\
11& 11 06  18.6& $-$77 57 53& 13.95&  0.89&  1.42& & \\
12& 11 06  27.7& $-$77 35 54& 13.26&  1.10&  1.65& & \\
13& 11 06  39.3& $-$77 36 04& 12.40&  1.23&  1.78& ISO-ChaI 79& 1 \\
14& 11 06  40.1& $-$76 48 35& 12.48&  0.94&  1.03&  & \\
15& 11 07  06.5& $-$76 37 17& 10.22&  1.50&  0.93& &  \\
16& 11 07  18.8& $-$77 31 56& 12.63&  1.93&  1.49& & \\
17& 11 07  27.9& $-$76 48 35& 13.91&  0.74&  0.38& & \\
18& 11 07  46.9& $-$76 15 17& 12.22&  0.71&  0.94& &  \\
19& 11 07  50.0& $-$77 38 07& 13.50&  1.55&  2.23& & \\
20& 11 08  20.6& $-$76 26 49& 13.80&  0.67&  0.62& & \\
21& 11 08  22.6& $-$76 49 19& 12.32&  1.12&  0.82& & \\
22& 11 08  41.9& $-$77 13 22& 13.78&  1.47&  0.83& & \\
23& 11 08  42.9& $-$77 10 08& 12.85&  0.95&  0.92& & \\
24& 11 08  44.5& $-$76 13 29& 12.92&  0.79&  0.56& & \\
25& 11 09  29.4& $-$76 34 46& 13.68&  0.95&  1.35&  NIR-14 & 2\\
26& 11 09  32.3& $-$76 30 17& 13.00&  0.94&  1.16& & \\
27& 11 09  33.9& $-$77 53 33& 13.90&  0.82&  1.19& & \\
28& 11 09  50.7& $-$78 04 19& 13.80&  0.77&  1.07& &  \\
29& 11 09  52.0& $-$76 39 12& 11.67&  0.69&  1.02&  ISO-ChaI 217 & 1\\
30& 11 09  53.4& $-$77 28 36& 12.48&  0.85&  1.24&  ISO-ChaI 220 & 1\\
31& 11 09  54.1& $-$76 31 11& 12.75&  0.98&  1.11&   ISO-ChaI 225 & 1\\
32& 11 10  01.6& $-$77 47 47& 13.97&  1.23&  1.46& & \\
33& 11 10  04.5& $-$77 48 22& 13.65&  1.21&  1.42& & \\
34& 11 10  06.2& $-$76 40 20& 13.85&  1.06&  1.53& &  \\
35& 11 10  16.8& $-$77 43 44& 13.01&  1.38&  1.28& & \\
36& 11 10  47.6& $-$76 32 38& 13.22&  0.84&  0.24& & \\
37& 11 10  49.4& $-$76 38 08& 13.87&  1.19&  0.76& & \\
38& 11 10  53.2& $-$76 16 53& 13.96&  0.78&  0.18& & \\
39& 11 10  56.9& $-$77 13 26& 13.93&  0.77&  1.03& & \\
40& 11 10  58.2& $-$76 17 56& 13.82&  0.77&  0.93& & \\
41& 11 11  08.9& $-$76 49 12& 13.21&  0.95&  1.54& & \\
42& 11 11  21.1& $-$78 05 20& 13.25&  0.69&  0.64& & \\
43& 11 11  27.8& $-$76 50 17& 13.48&  0.90&  1.09& & \\
44& 11 11  35.7& $-$76 53 26& 13.19&  1.16&  1.79& & \\
45& 11 11  42.6& $-$78 05 21& 13.88&  0.66&  0.53& &  \\
46& 11 11  48.7& $-$76 12 21& 13.83&  0.75&  0.49& & \\
47& 11 11  54.5& $-$76 58 29& 13.71&  1.29&  0.98& &  \\
48& 11 11  57.5& $-$77 21 57& 13.72&  1.38&  1.13& &  \\
49& 11 12  03.7& $-$76 51 33& 13.51&  1.27&  1.82& & \\
50& 11 13  25.3& $-$77 00 23& 11.55&  0.98&  0.67& &  \\
51& 11 13  26.0& $-$77 00 32& 12.40&  1.20&  0.28& &  \\
52& 11 13  42.7& $-$77 55 30& 13.71&  0.45&  0.25& & \\
53& 11 14  31.4& $-$78 06 44& 13.03&  0.98&  0.75& & \\
54& 11 14  23.6& $-$77 56 12& 13.73&  0.84&  0.80& & \\
55& 11 16  13.2& $-$77 25 15& 11.00&  0.85&  0.21& & \\
56& 11 18  09.0& $-$78 14 11& 13.71&  1.03&  0.72& & \\
\enddata

\tablecomments{Units of right ascension are hours, minutes, and seconds, and
units of declination are degrees, arcminutes, and arcseconds.}

\tablerefs{(1) \citet{per00}, (2) \citet{oas99}}

\end{deluxetable}

\begin{deluxetable}{cccccccl}
\tablecaption{Candidate Pre-Main Sequence Stars with $K$ $>$ 14.0. \label{tbl-4}}
\tablewidth{0pt}
\tablehead{
\colhead{Star}& \colhead{$\alpha$(2000.0)} & \colhead{$\delta$(2000.0)} & \colhead{$K$} &
\colhead{$H-K$} & \colhead{$J-H$} & \colhead{Id.}  & \colhead{Ref.} }
\startdata
57& 10 40 47.6& $-$77 36 49& 15.12& 0.95& 0.73& & \\
58& 11 04 17.1& $-$77 25 49& 15.34& 0.86& 0.65& & \\
59& 11 04 27.0& $-$77 28 02& 14.87& 1.14& 1.02& & \\
60& 11 04 30.5& $-$77 34 41& 14.33& 0.96& 0.89& & \\
61& 11 04 32.0& $-$77 54 09& 15.27& 1.02& 0.88& & \\
62& 11 04 35.8& $-$77 29 09& 15.34& 0.99& 0.94& & \\
63& 11 04 36.2& $-$77 50 12& 14.76& 1.21& 1.26& & \\
64& 11 04 38.7& $-$77 24 21& 14.49& 1.27& 0.87& & \\
65& 11 04 39.1& $-$77 40 05& 14.39& 1.56& 1.18& & \\
66& 11 04 48.4& $-$77 49 20& 15.45& 1.69& 0.85& & \\
67& 11 05 37.0& $-$76 51 18& 15.33& 1.48& 1.04& & \\
68& 11 05 41.2& $-$76 38 38& 15.12& 0.91& 0.58& & \\
69& 11 05 42.6& $-$77 26 30& 14.33& 2.09& 2.67& & \\
70& 11 05 43.0& $-$77 31 58& 14.43& 1.46& 1.16& & \\
71& 11 06 33.5& $-$77 52 26& 14.79& 0.98& 0.74& & \\
72& 11 06 35.3& $-$77 21 10& 15.94& 1.20& 0.95& & \\
73& 11 06 49.4& $-$77 34 37& 14.99& 1.51& 1.78& & \\
74& 11 06 51.0& $-$77 11 33& 15.08& 1.16& 0.64& & \\
75& 11 06 54.5& $-$77 25 54& 15.30& 1.36& 0.70& & \\
76& 11 06 59.6& $-$77 18 29& 14.90& 1.59& 1.52& & \\
77& 11 07 59.0& $-$78 00 13& 15.05& 1.41& 0.84& & \\
78& 11 08 27.3& $-$77 47 20& 14.75& 1.44& 0.93& & \\
79& 11 08 51.6& $-$77 02 30& 14.82& 1.75& 0.65& & \\
80& 11 08 51.5& $-$77 06 44& 15.10& 0.87& 0.64& & \\
81& 11 09 08.3& $-$78 14 25& 15.09& 1.31& 0.69& & \\
82& 11 09 18.8& $-$77 39 37& 15.10& 1.26& 1.10& & \\
83& 11 09 24.1& $-$76 34 55& 14.87& 1.46& 1.56& NIR-11 & 2 \\
84& 11 09 24.7& $-$78 13 54& 15.69& 0.86& 0.78& & \\
85& 11 09 52.7& $-$77 49 08& 14.87& 1.80& 1.37& & \\
86& 11 10 01.7& $-$77 19 20& 15.35& 1.05& 1.00& &\\
87& 11 10 04.3& $-$77 27 15& 15.34& 2.01& 0.66& & \\
88& 11 10 17.7& $-$77 22 18& 15.69& 1.31& 1.03& & \\
89& 11 10 19.6& $-$77 11 20& 14.57& 0.94& 0.89& & \\
90& 11 10 24.6& $-$77 23 06& 15.58& 1.93& 0.73& & \\
91& 11 10 26.2& $-$76 44 09& 15.99& 1.18& 0.60& & \\
92& 11 10 29.4& $-$76 31 12& 15.36& 1.37& 0.58& & \\
93& 11 10 29.8& $-$77 21 12& 15.11& 1.43& 0.64& & \\
94& 11 10 37.9& $-$77 01 39& 15.29& 1.21& 0.75& & \\
95& 11 10 39.2& $-$77 02 48& 15.44& 1.45& 1.10& & \\
96& 11 10 46.1& $-$77 17 07& 14.47& 0.99& 0.88& & \\
97\tablenotemark{a} & 11 11 06.3& $-$78 03 57& 14.74& 1.00& 0.98& & \\
98& 11 11 11.5& $-$76 44 37& 15.02& 1.24& 1.21& & \\
99& 11 11 12.3& $-$76 47 51& 15.28& 1.01& 0.99& & \\
100\tablenotemark{a} & 11 11 15.6& $-$78 13 22& 14.71& 1.28& 0.87&& \\
101& 11 11 21.3& $-$77 24 15& 15.02& 0.97& 0.80& & \\
102& 11 11 28.6& $-$78 05 28& 15.02& 1.61& 1.03& & \\
103& 11 11 30.2& $-$78 10 10& 15.06& 0.89& 0.72& & \\
104& 11 11 34.3& $-$76 47 35& 15.34& 0.82& 0.67& & \\
105& 11 11 47.3& $-$77 08 12& 14.50& 1.40& 1.00& & \\
106& 11 11 48.1& $-$77 25 46& 15.06& 0.95& 0.81& & \\
107& 11 12 05.1& $-$77 29 58& 15.22& 1.17& 1.16& & \\
108& 11 12 13.1& $-$77 22 37& 14.47& 1.07& 1.02& & \\
109& 11 13 02.6& $-$77 22 08& 14.72& 1.13& 0.74& & \\
110& 11 13 23.8& $-$77 25 28& 15.64& 1.55& 1.55& & \\
111& 11 13 53.1& $-$77 37 01& 14.62& 1.93& 1.91& & \\
112\tablenotemark{a} & 11 13 56.0& $-$77 06 35& 14.28& 1.13& 0.72& & \\
113& 11 14 17.3& $-$76 48 18& 15.11& 0.95& 0.86& & \\
114& 11 15 00.3& $-$78 03 13& 15.04& 0.96& 0.89& & \\
115& 11 16 30.3& $-$78 05 08& 15.30& 1.12& 0.77& & \\
116& 11 16 46.0& $-$77 37 40& 15.06& 0.89& 0.74& & \\
117& 11 17 36.8& $-$77 45 35& 15.18& 1.03& 0.68& & \\
118& 11 18 02.2& $-$77 41 30& 15.61& 0.86& 0.58& & \\
\enddata

\tablecomments{Units of right ascension are hours, minutes, and seconds, and
units of declination are degrees, arcminutes, and arcseconds.}

\tablenotetext{a}{Star with fuzzy image at $K$.}

\tablerefs{(1) \citet{per00}, (2) \citet{oas99}}

\end{deluxetable}



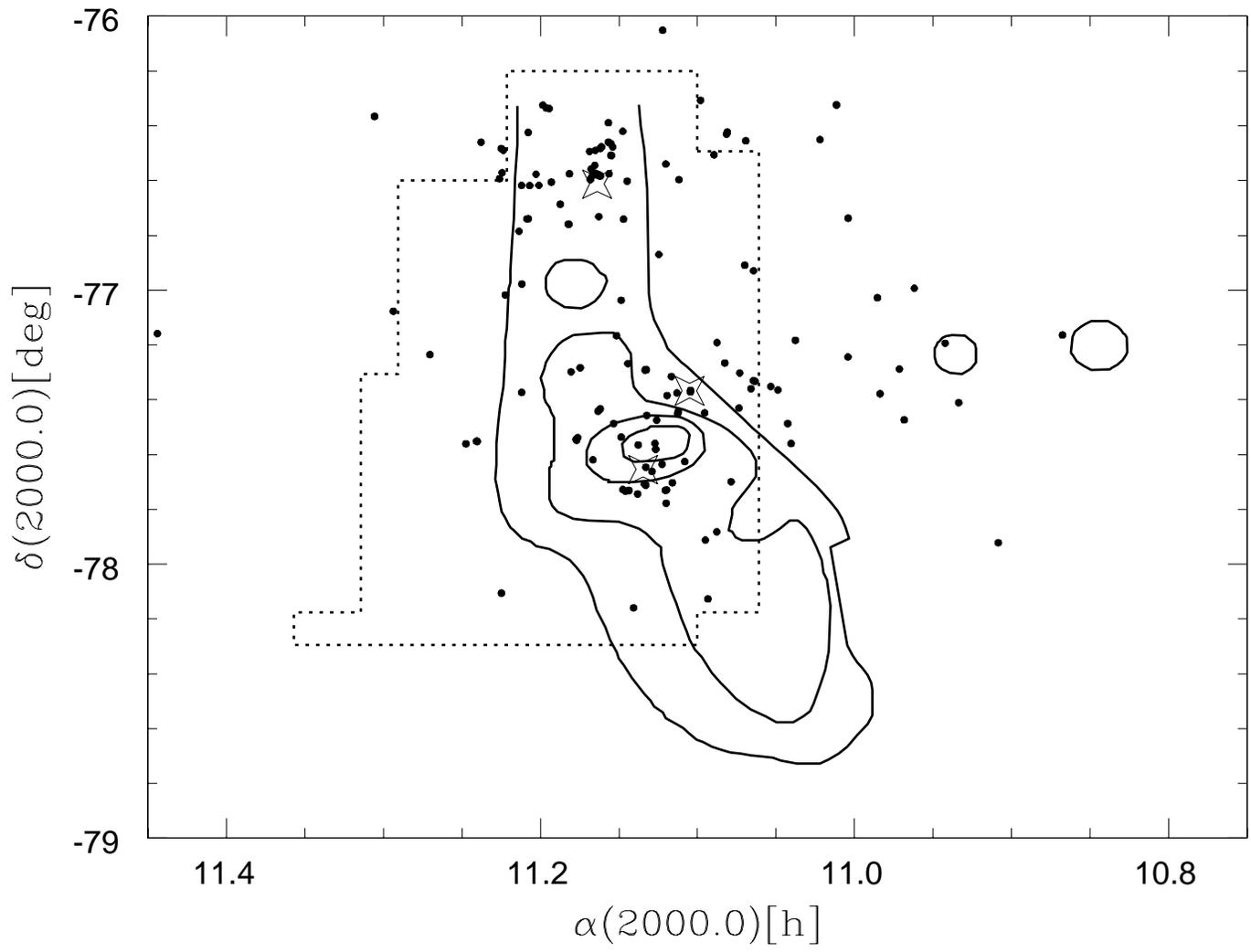
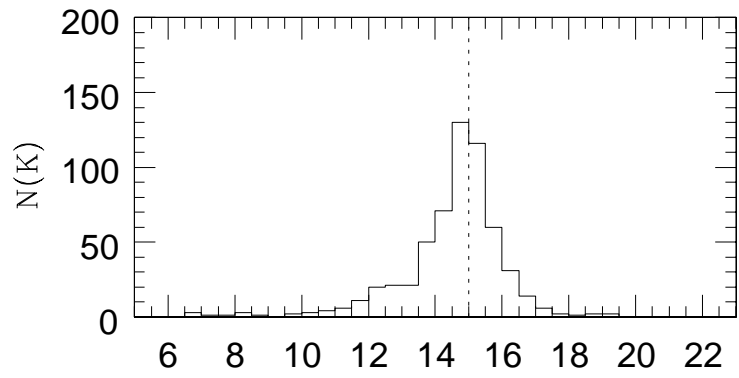
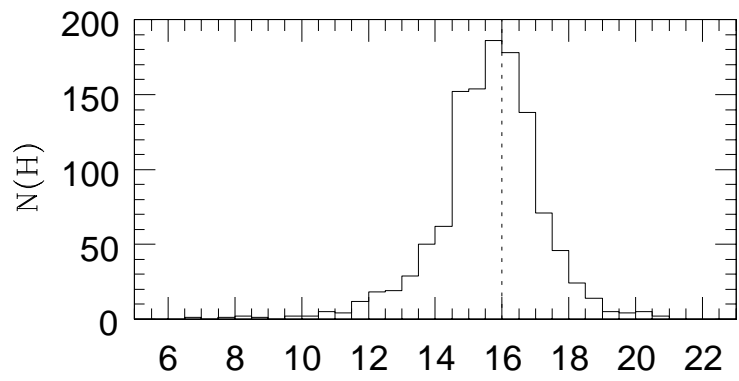
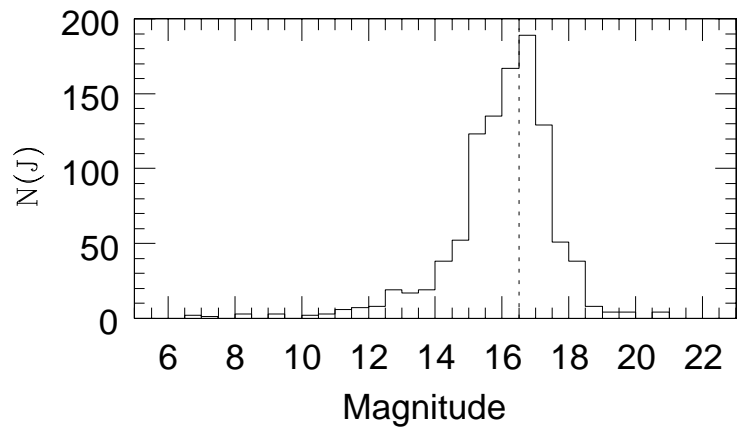
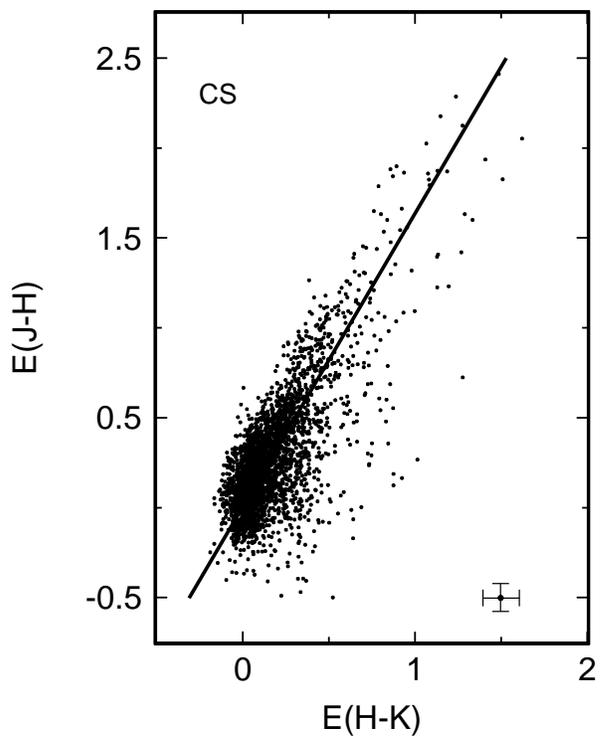
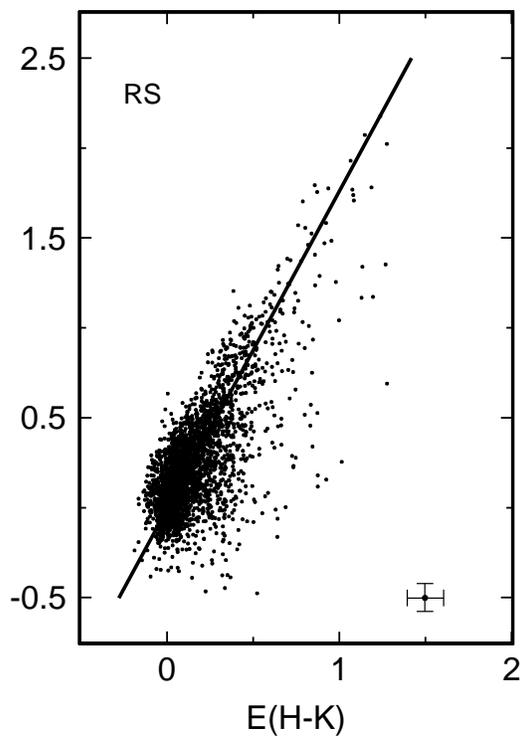
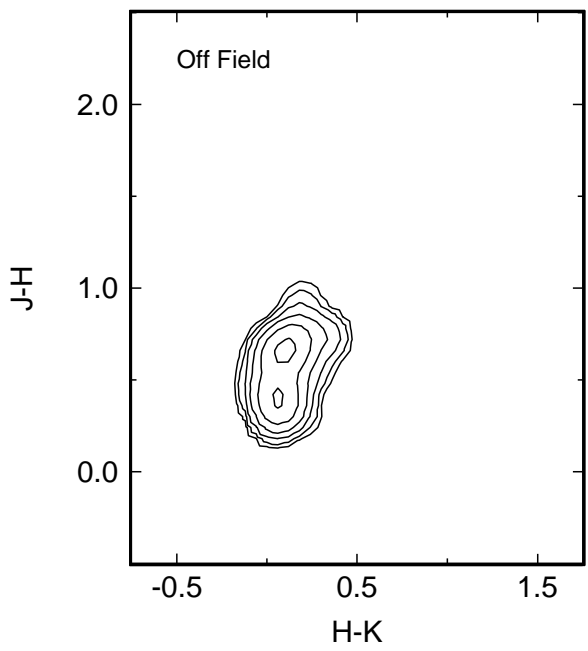
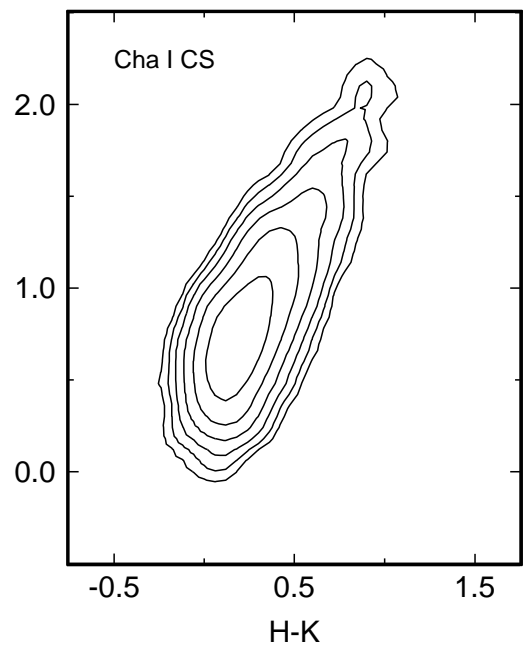
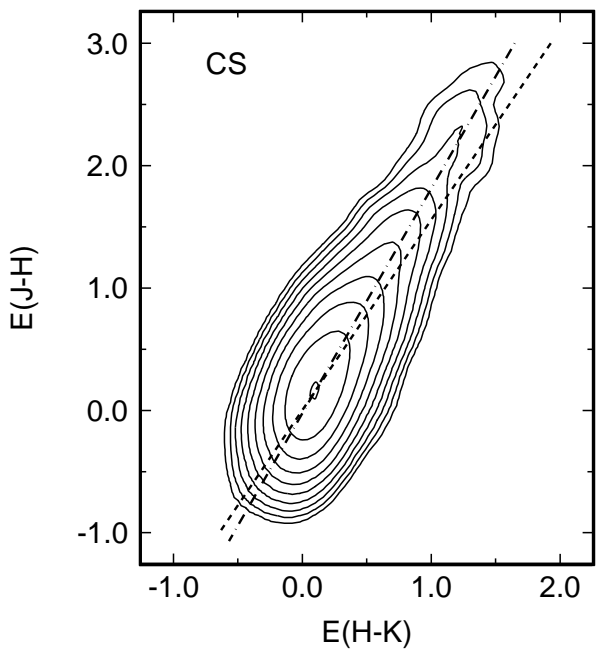
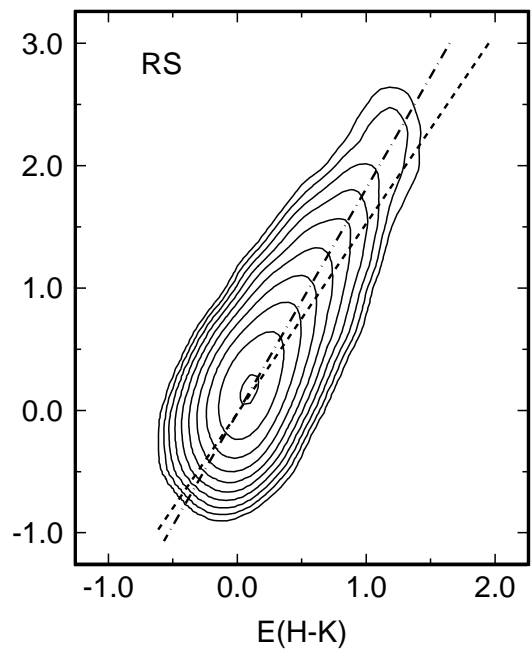
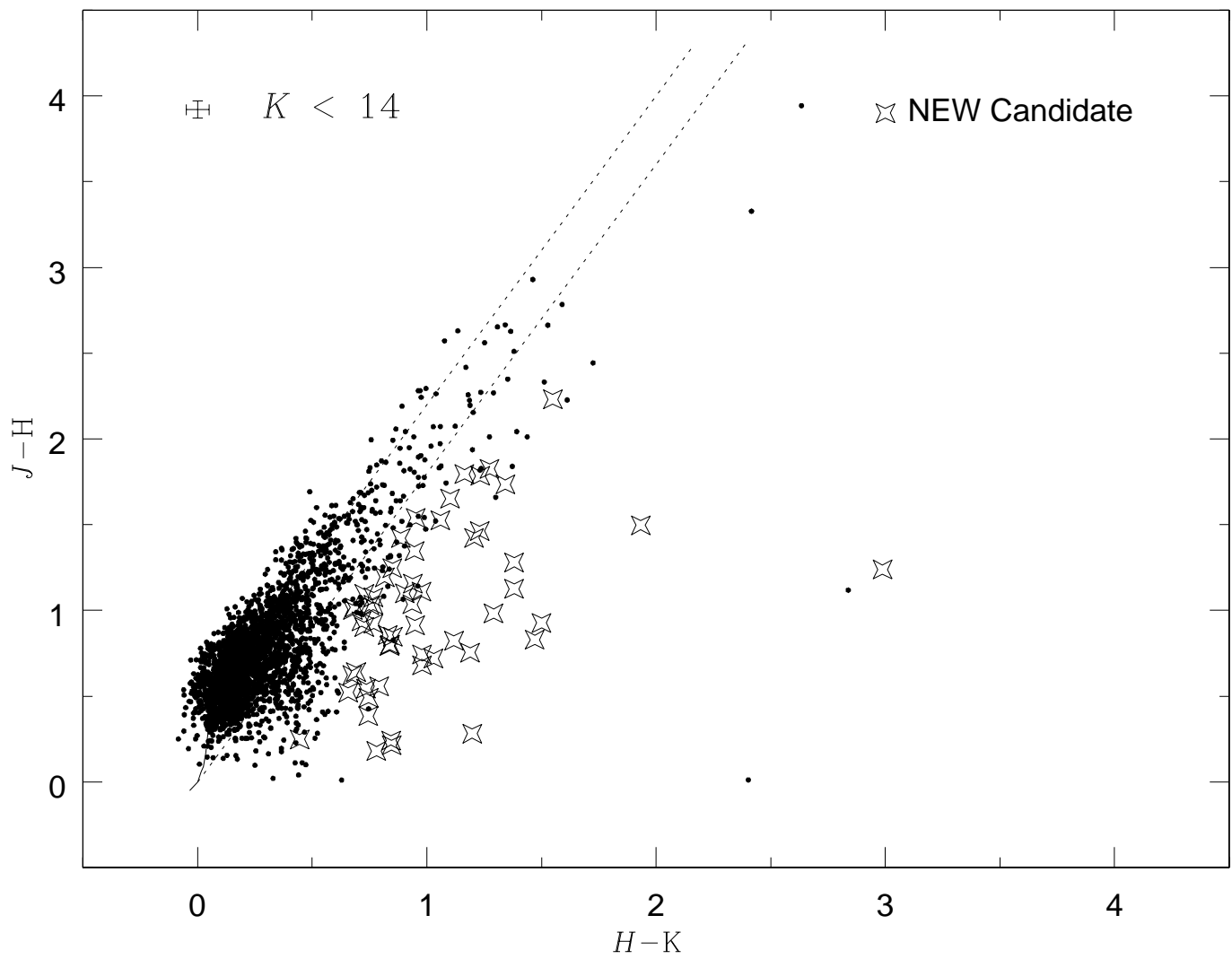
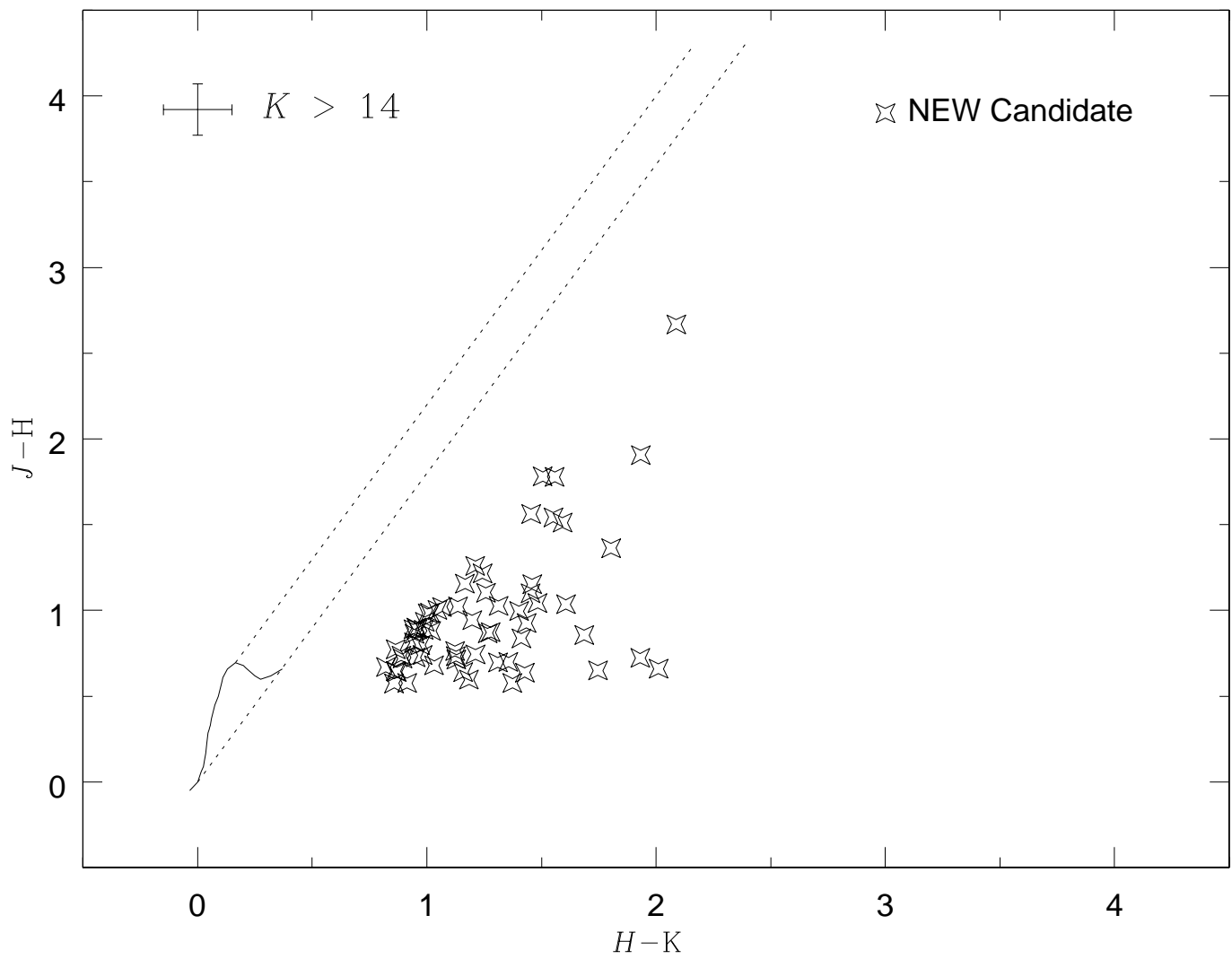
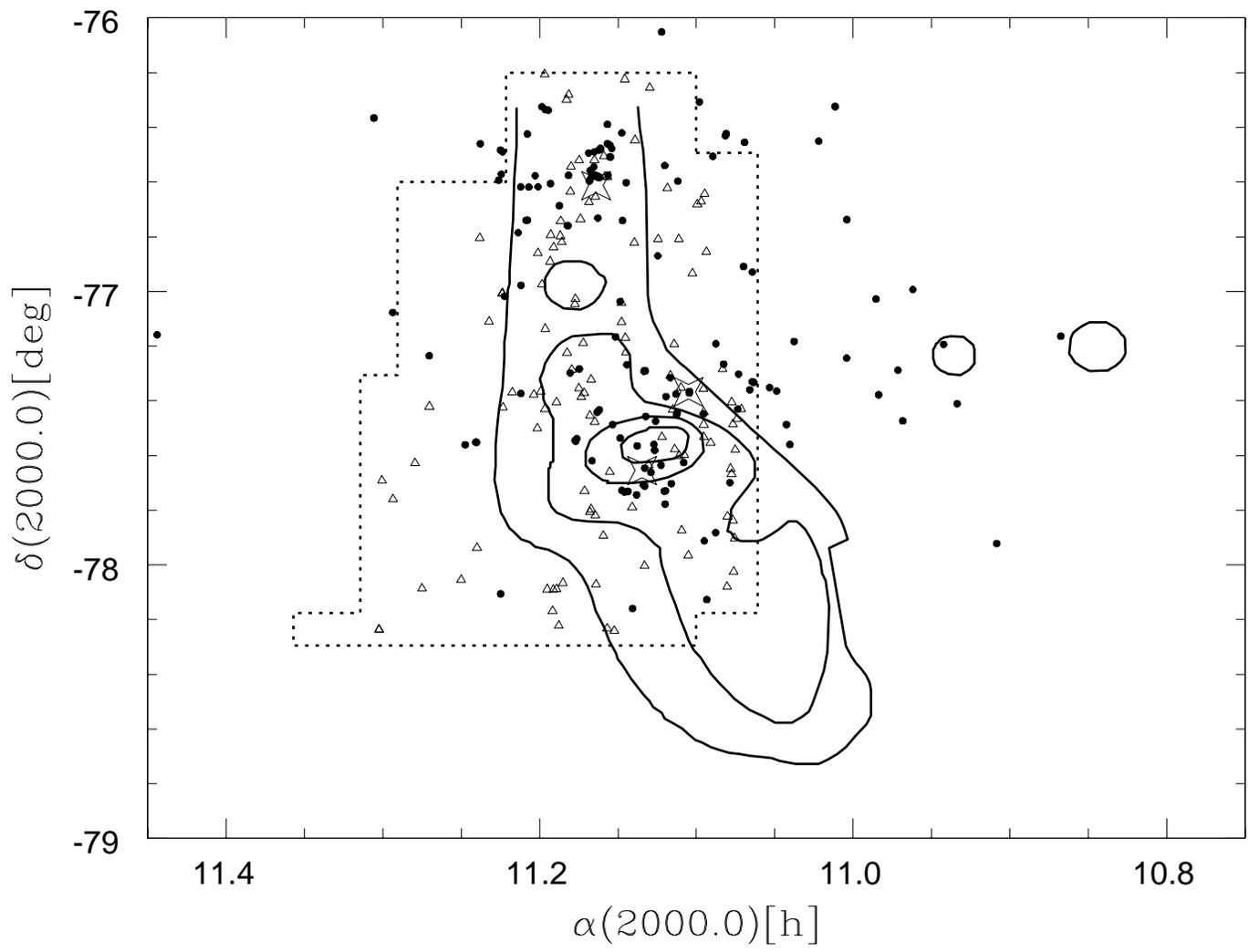
\end{document}